\documentclass[showpacs,preprintnumbers,preprint,aps,amsmath,amssymb]{revtex4}
\usepackage{epsfig}
\def\lsim{\:\raisebox{-1.1ex}{$\stackrel{\textstyle<}{\sim}$}\:}
\def\gsim{\:\raisebox{-1.1ex}{$\stackrel{\textstyle>}{\sim}$}\:}
\def\10{$SO(10)$}
\def\21{SU(2) $\otimes$ U(1) }

\def\422{$SU(4) \otimes SU(2) \otimes SU(2)$}
\def\321{SU(3) $\otimes$ SU(2) $\otimes$ U(1)}
\def\lsim{\raise0.3ex\hbox{$\;<$\kern-0.75em\raise-1.1ex\hbox{$\sim\;$}}}
\def\gsim{\raise0.3ex\hbox{$\;>$\kern-0.75em\raise-1.1ex\hbox{$\sim\;$}}}
\def\vev#1{\left\langle #1\right\rangle}

\newcommand{\ba}{\begin{array}}
\newcommand{\ea}{\end{array}}
\newcommand{\be}{\begin{equation}}
\newcommand{\ee}{\end{equation}}
\newcommand{\beqa}{\begin{eqnarray}}
\newcommand{\eeqa}{\end{eqnarray}}
\relax
\def\321{$SU(3)\times SU(2)\times U(1)$}

\newcommand{\bqbqbar}{$B^0_q-\bar{B}^0_q$}

\begin{document}
\bigskip
\title[]{Minimal Flavour Violations and Tree Level FCNC} 

\author{ Anjan S.  Joshipura\footnote{anjan@prl.res.in} and Bhavik P.  
Kodrani\footnote{bhavik@prl.res.in}} \affiliation{ Physical Research
Laboratory, Navarangpura, Ahmedabad 380 009, India \vskip 2.0truecm}

\begin{abstract} 
\vskip 1.0 truecm 

Consequences of a specific class of two Higgs doublet models in which the
Higgs induced tree level flavour changing neutral currents (FCNC) display
minimal flavour violation (MFV) are considered.  These FCNC are fixed in
terms of the CKM matrix elements and the down quark masses. The minimal
model in this category with only two Higgs doublets has no extra CP
violating phases but such a phase can be induced by adding a complex
singlet.  Many of the theoretical predictions are similar to other MFV
scenario. The FCNC contribute significantly to $B$ meson mixing and CP
violation. Detailed numerical analysis to determine the allowed Higgs
contributions to neutral meson mixings and the CKM parameters
$\bar{\rho},\bar{\eta}$ in their presence is presented. The Higgs induced
phase in the $B^0_{d,s}-\bar{B}^0_{d,s}$ transition amplitude
$M_{12}^{d,s}$ is predicted to be equal for the $B_d$ and the $B_s$
systems. There is a strong correlation between phases in $M_{12}^{d,s}$
and $|V_{ub}|$. A measurable CP violating phase $\phi_s=-0.18\pm 0.08$ is
predicted on the basis of the observed phase $\phi_d$ in the $B_d$ system
if $|V_{ub}|$ is large and close to its value determined from the
inclusive b decays. 
\end{abstract} 

\maketitle
 
%%%%%%%%%%%%%%%%%%%%%%%%%%%%%%%%%%%%%%%%%%%%%%%%%%%%%%%
\section{Introduction}
%%%%%%%%%%%%%%%%%%%%%%%%%%%%%%%%%%%%%%%%%%%%%%%%%%%%%%%
The Cabibbo Kobayashi Maskawa (CKM) matrix $V$ provides a unique
source of flavour and CP violations in the standard model (SM). It leads
to flavour changing neutral currents (FCNC) at the one loop level.
$K$ and $B$ meson decays and mixing have provided stringent
tests of these FCNC induced processes and the SM predictions have been 
verified  with  some hints for 
possible new physics contributions \cite{np1,bf,np2}. Any 
new source 
of flavour violations 
resulting from the well-motivated extensions of the SM ($e.~g.$ 
supersymmetry) is now constrained  to be small \cite{utfit1,utfit2,ckmfit}.

Several extensions of the SM share an important property termed as 
minimal
flavour violation (MFV) \cite{effectivemfv,mfv}.  
According to this, all
flavour and CP violations are determined by  the CKM matrix even when the 
SM is
extended to include other flavour violating interactions. In the extreme 
case
(termed as the constrained MFV \cite{cmfv}) the 
operators responsible for the flavour violations are also the same as
in the SM. In more general situations, MFV models contain more operators 
with coefficients 
determined in terms of the elements of $V$. Some scenarios 
\cite{nmfv} termed as
the next to  minimal flavour violation (NMFV)  contain new phases 
not present in $V$.

A simple example of MFV is provided by a  two Higgs doublet model with 
natural flavour 
conservation (NFC) \cite{nfc}. The discrete symmetry conventionally 
imposed to obtain
NFC prevents any CP violation coming from the Higgs potential and the CKM 
matrix provides a unique source of CP and flavour violations in these 
models. The MFV in these models can be explicitly seen by 
considering the  
$B^0_q$-$\bar{B}^0_q$ ($q=d,s$)  transition amplitude $M_{12}^q$ as an 
example. The charged Higgs 
boson in the model gives  additional contributions 
to the SM amplitude and
the dominant top quark dependent part can be written \cite{franzini} as
\be \label{k+}
M_{12}^q= \frac{G_F^2 M_W^2 M_{B_q} B_{q}f_{B_q}^2\eta_B(x_t) 
(V_{tb}V_{tq}^*)^2}{12 
\pi^2}(1+\kappa^+_H) ~, \ee
where
\beqa 
\kappa^+_H&\equiv& \frac{1}{4 
S_0(x_t)}\frac{\eta_B(x_t,y_t)}{\eta_B(x_t)}(\cot^4 \theta 
S_{HH}(y_t)+\cot^2\theta
S_{HW}(x_t,y_t))~, \nonumber \\
&\approx&\frac{\eta_B(x_t,y_t)}{\eta_B(x_t)}(0.12 \cot^4 \theta+0.53 
\cot^2\theta) ~, \eeqa
where $\eta_B$ are the  QCD corrections \cite{nlo,2Hdm1}, $\tan \theta$ is 
the 
ratio of the 
Higgs vacuum expectation values and 
$x_t=\frac{m_t^2}{M_W^2},y_t=\frac{m_t^2}{M_H^{+^2}}$. The functions 
appearing 
above can be found for example in \cite{2Hdm1,2Hdm2} and the last line 
corresponds 
to the obtained numerical values in case of the charged Higgs mass  
$M_{H^+}=200$ GeV.  Flavour and  CP violation  are still
governed by the same combinations of the CKM matrix elements that appear
in the SM box diagram.
The only effect of the charged Higgs boson is an additional contribution 
to the function $S_0(x_t)$. The same 
happens in case of other observables and one can parameterize all the
FCNC induced processes in terms of seven  
independent functions in MFV models\cite{mfv}.

Two Higgs doublet models (2HDM) with NFC lead to  MFV but they do 
not represent the 
most generic 
possibilities.
More general 2HDM will contain additional sources of CP and flavour 
violation through the presence of FCNC. The principle of NFC now appears
to conflict \cite{mb} with the idea of the spontaneous CP violation 
(SPCV) at 
low energy and both 
cannot coexist together. Indeed, if NFC and the spontaneous CP violation 
are 
simultaneously present in  multi-Higgs doublet models then the CKM matrix 
is 
implied to 
be real \cite{brancoreal}. In contrast, the detailed model independent 
fits to experimental 
data
require the Wolfenstein parameter $\bar{\eta}=0.386\pm 0.035$ according to 
the latest 
fits by the UTfit collaboration \cite{utfit2}. Thus the CKM matrix is 
proven to be complex
under very general assumptions \cite{brancockm}. Attractive idea of 
low energy 
SPCV can only be realized by admitting the tree level FCNC \cite{bhavik}. 
Independent of this, the 2HDM without NFC become 
phenomenologically interesting if there is a natural mechanism to suppress
FCNC. The phenomenology of such 
models has been studied in variety of context \cite{fcncm}.

This paper is devoted to discussion of models in which FCNC are naturally 
suppressed and show strong hierarchy \cite{asj1,asj2,bgl}. Specifically, 
the FCNC couplings 
$F_{ij}^{d}$ between the $i$ and the $j$ generations obey
\be \label{hier}
|F_{12}^{d}|<|F_{13}^{d}|,|F_{23}^{d}| \ee
automatically suppressing the flavour violations in the $K$ sector
relative to  $B$ mesons. 
A specific sub-class of these models has the remarkable property that 
the FCNC couplings are determined completely in terms of  the CKM matrix 
and the quark masses \cite{bgl}. These models 
therefore provide yet another example of MFV in spite of the 
presence of FCNC. The models to be discussed were presented long ago 
\cite{asj1,asj2,bgl}
and 
the aim of the present paper is to update constraints on them in view of 
the
substantial  experimental information that has become available from the 
Tevatron and $B$ factories.

The next section introduces the class of models we discuss and presents 
the structure of the FCNC couplings. Section (III) is devoted to the 
analytic and numerical studies of the consequences assuming that either
the charged Higgs or a neutral Higgs dominates the $P^0$-$\bar{P}^0$ 
($P=K,B_d,B_s$) 
mixing. The last section summarizes the salient features of the paper.

\section{Model and the structure of FCNC}
Consider the \21 model with two Higgs doublets $\phi_{a}$, 
($a=1,2$) and the following Yukawa couplings:
\begin{equation}
\label{yukawa}
 -{\cal L}= \bar{Q'}_L\Gamma^d_a\phi_a
d_R'+\bar{Q'}_L\Gamma^u_a\tilde{\phi}_a u_R'+ {\rm H.c.}~.
\end{equation}
$Q^{'}_{iL}$  ($i=1,2,3$) represent three generations of weak doublets and 
$u^{'}_{iR},d^{'}_{iR}$ are the 
corresponding singlets. Let us consider a class of models \cite{asj1} 
represented by
a specific choice of the matrices $\Gamma_a^d$ and their permutations:
\be
\label{gammad}
\ba{cc}
\Gamma^d_1=\left( \ba{ccc}
x&x&x\\
x&x&x\\
0&0&0\\ \ea \right)
~&~;~~
\Gamma^d_2=\left( \ba{ccc}
0&0&0\\
0&0&0\\
x&x&x\\ \ea \right)
\ea~,\ee
where $x$ represents an entry which is allowed to be non-zero. 
We do not impose CP on eq.(\ref{yukawa}) allowing elements in 
$\Gamma_{1,2}^d$ to be complex. The above
forms of $\Gamma_a^d$ are technically natural as they follow from 
imposition
of discrete symmetries on eq.(\ref{yukawa}), the simplest being a $Z_2$
symmetry under which only $Q^{'}_{3L}$ and $\phi_2$ change sign. 

The down quark mass matrix $M_d$ follows from eq.(\ref{gammad}) when 
the Higgs 
fields obtain their vacuum expectation values (vev): $\vev{\phi_1^0}=v_1$ 
and
$\vev{\phi_2^0}=v_2 e^{i\alpha}$. Let $V_{d L,R}$ be the unitary matrices 
connecting the mass (unprimed) and the weak basis $d^{'}_{L,R}=V^d_{L,R} 
d_{L,R}$. Then
\be \label{diagmd} 
V_{dL}^\dagger M_d V_{dR}=D_d ~, \ee
$D_d$ being a diagonal matrix of the down quark masses $m_i$. $M_d$ 
obtains contributions from two different Higgs fields leading to the Higgs 
induced FCNC in the down quark sector. 
Eqs. 
(\ref{yukawa}-\ref{diagmd}) are used to obtain:
\be \label{fcnc}
-{\cal L}_{FCNC}=\frac{(2 \sqrt{2} G_F)^{1/2}}{\sin\theta\cos\theta}
F^d_{ij}\bar{d}_{iL} d_{jR}\phi^0
+{\rm H.C.} ~, \ee
where $\tan\theta=\frac{v_2}{v_1}$ and 
\be \label{phih}\phi^0 \equiv 
\cos\theta~\phi^0_2~e^{-i\alpha}-\sin\theta~\phi^0_1~ \ee
is a specific combination of $\phi_{1,2}$ with zero vev. The orthogonal 
combination plays the role of the standard model Higgs. The strength 
of FCNC current
is determined in the fermion mass basis by \cite{asj1}:
\be \label{fij1}
F_{ij}^d\equiv (V^{d\dagger}_L\Gamma_2^d v_2 e^{i\alpha}
V^{d}_R)_{ij}=(V^*_{dL})_{3i} (V_{dL})_{3j} m_j~,\ee
Note that the the specific 
texture of 
$\Gamma_{1,2}^d$ allowed us to express $F_{ij}^d$ in terms of the left 
handed 
mixing and the down quark masses $m_j$ and the dependence on the 
unphysical $V_{dR}$ 
disappeared. The $F_{ij}^d$ depend on the left-handed mixing matrix 
$V_{dL}$ which is a priori  unknown but would be correlated to the CKM 
matrix. One observes that 
\begin{itemize}  
\item  independent of the values of elements of $V_{dL}$, the $F_{ij}^d$
display hierarchy given in eq.(\ref{hier}).
\item  all the FCNC couplings are suppressed if the off-diagonal elements 
of $V_{dL}$ are smaller than the diagonal ones. The model in this sense 
illustrates the principle of near flavour conservation  
\cite{hall}.  This is a generic 
possibility in view of the strong mass hierarchy among  quarks unless 
there are some special symmetries.
\item $F_{ij}^d$ can be determined in terms of the CKM matrix elements for 
a 
specific structure of $M_u$ \cite{bgl} given as follows: 
\be \label{mu}
M_u=\left( \ba{ccc} x&x&0\\
                    x&x&0\\
                    0&0&x\\ \ea \right) ~. \ee
The above postulated structures of $M_{u,d}$ follow from  discrete 
symmetries \cite{bgl} rather than being ad-hoc. Particular example can be:
\beqa \label{symmetry} 
(Q^{'}_{1,2L},\phi_1)\rightarrow \omega ( Q^{'}_{1,2L},\phi_1)
&,& u^{'}_{1,2 R}\rightarrow \omega^2 u^{'}_{1,2 R} ~.
\eeqa
Here $\omega,\omega^2\not = 1$ are complex numbers. The fields not shown 
above remain unchanged under the symmetry.

The particular form of $M_u$ as given above implies 
that $(V_{dL})_{3i}=V_{3i}$ as a result of which the $F^d_{ij}$ in 
eq.(\ref{fij1}) are completely determined in terms 
of the CKM matrix 
$V$.
\be \label{fij2} F_{ij}^d=V^*_{3i}V_{3j} m_j ~. \ee
As a consequence of eq.(\ref{symmetry}), $(M_u)_{33}$ gets contribution 
from $\phi_2$ while  the first two generations from $\phi_1$ with no 
mixing 
with the third one. As a result, there are no FCNC in the up quark 
sector while they are determined as in eq.(\ref{fij2}) in the down quark 
sector.
\end{itemize}

The
tree level couplings of the charged Higgs $H^+\equiv
\cos \theta \phi_2^+-\sin\theta e^{i \alpha}  
\phi_1^+$ can be read off from 
eq.(\ref{yukawa}) 
and 
are given by
\be \label{ffh+}
(2\sqrt{2} G_F)^{1/2}H^+ \left\{\bar{u}_R\hat{D}_u V d_L+ \bar{u}_L(VD_d 
\tan \theta   
-\frac{1}{s_\theta c_\theta}VF^d)d_R  
\right\}+{\rm H.C.} ~,\ee
where $\hat{D}_u\equiv {\rm diag.}(-m_u \tan\theta, 
-m_c\tan\theta,m_t\cot\theta)$. 

It follows from eqs.(\ref{fcnc},\ref{fij2},\ref{ffh+}) that 
all the Higgs fermion couplings are determined by the CKM 
matrix $V$ giving rise to MFV. There can however be an additional source 
of CP violation in the model. This can arise if 
the scalar-pseudoscalar mixing contains a phase. As noted in \cite{bgl}, 
the discrete symmetry of eq.(\ref{symmetry}) prevents this mixing in the
Higgs potential even if one allows for explicit CP violation and a 
bilinear soft symmetry breaking 
term $\mu (\phi_1^\dagger\phi_2)+{\rm H.c.}$. Thus the minimal version of 
the model corresponds to the MFV scenario with no other CP violating 
phases present. 
CP violation in Higgs mixing can however be induced by adding a 
complex 
Higgs singlet field \cite{bgl,asj}. In this case,
there would be an additional phase which mixes the real and the imaginary
parts of the Higgs $\phi^0$ defined in eq.(\ref{phih}). We will admit such 
a phase in our discussion.

There is an important quantitative difference between the present scenario 
and the general 
MFV analysis following from the effective field theory approach 
\cite{effectivemfv}.
There the effective dominant FCNC couplings between down quarks are given 
by 
$$ (\lambda_{FC})_{ij}\approx \lambda_t^2 V^*_{3i} V_{3j}~,$$
where $\lambda_t$ denotes the top Yukawa coupling. The same factor 
controls the loop induced contributions here but the tree 
level  flavour violations are given by eq. (\ref{fij2}) which  contains 
the same elements of  $V$ but involves the down quark masses linearly. Its 
contribution 
is still important or dominates over the top quark dependent terms because 
of
its presence at the tree level.
 
One could consider  variants of the above textures and symmetry obtained 
by permutations of flavour indices. These variants lead to different amount 
of FCNC. Labeling these variants by $a$, one has three models 
\cite{bgl} with
$F_{ij}^d(a)=V^*_{ai}V_{aj}m_j~, (a=1,2,3)$. Alternatively, one could also 
consider
equivalent models in which FCNC in the d quarks are absent while in the up 
quark sector they would be related to the CKM matrix elements and the up 
quark masses. The case $a=3$ is special. It leads to the maximum 
suppression of FCNC in the 12 sector.   
We will mainly consider phenomenological implication of that case. 

\section{Experimental constraints and their implications  }
\subsection{Basic Results}
The strongest constraints on the model come from the $P^0-\bar{P}^0$ 
($P=K,B_d,B_s$) mixing. In addition to the SM contribution, two 
other 
sources namely,  the charged Higgs induced box diagrams
and the  neutral Higgs $\phi^0$ induced tree diagram contribute to
this mixing. 

The charged Higgs leads to new box diagrams which follow from 
eq.(\ref{ffh+}). The last two terms of this equation  are suppressed by 
the down quark masses (for modest 
$\tan\theta$) and the dominant contribution comes from the top quark. This
term  and hence the charged Higgs contributions remain the 
same as in 2HDM with NFC \cite{franzini}. The 
contribution to
the \bqbqbar   ~mixing is already given in eq.(\ref{k+}).
The contribution to $\epsilon$ is given \cite{2Hdm2} by
\be \label{epsh+}
\epsilon^{H^+}=\frac{G_F^2 M_W^2 f_K^2 m_K B_K A^2 
\lambda^6\bar{\eta}}{6\sqrt{2}\pi^2\Delta 
m_K}\left(f_1^H+f_2^H A^2 
\lambda^4(1-\bar{\rho})\right)~,\ee
where functions $f_{1,2}^H$ can be read-off from expressions given in 
\cite{2Hdm2}. 
$\lambda, \bar{\eta}\equiv \eta (1-\frac{\lambda^2}{2}),\bar{\rho}\equiv 
\rho(1-\frac{\lambda^2}{2})$ and $A$ are the Wolfenstein parameters.
Contribution of $f_1^H$ to $\epsilon$ is practically negligible while the
$f_2^H$ can compete with the corresponding term in the SM expression
\be \label{epssm}
\epsilon^{SM}=\frac{G_F^2 M_W^2 f_K^2 m_K B_K A^2 
\lambda^6\bar{\eta}}{6\sqrt{2}\pi^2\Delta 
m_K}(f_1(x_t)+f_2(x_t) A^2 
\lambda^4(1-\bar{\rho}))\ee
for moderate values of $\tan\theta$.

The neutral Higgs contributions to the above observables follow from 
eqs.(\ref{fcnc}) and (\ref{fij2}). Define
$$ \phi^0\equiv\frac{R+i I}{\sqrt{2}}=\left(\frac{O_{R\alpha}+i 
O_{I\alpha}}{\sqrt{2}}\right)H_\alpha^0\equiv |C_\alpha| e^{i\eta_\alpha} 
H_\alpha^0~, $$
where $H_\alpha^0$ denote the mass eigenstates with masses $M_\alpha$.
$\alpha=1,2,3$ for the 2HDM while $\alpha=1,...5$ in the presence of a 
complex singlet introduced to induce the scalar-pseudo scalar mixing
leading to phases $\eta_\alpha$ in the Higgs mixing  $C_\alpha$. 
$O_{R\alpha,I\alpha}$ are elements of the mixing matrix.  
Using this 
definition and eq.(\ref{fij2}) the neutral Higgs contribution to $
M_{12}^q$ 
can be written as
\be\label{m12qh0}
(M_{12}^{q})^{H^0}=\frac{5 \sqrt{2}G_F m_b^2m_{B_q}f_{B_q}^2 B_{2q}}{12 
\sin^2 
2\theta M_\alpha^2}\left(\frac{m_{B_q}}{m_b+m_q}\right)^2C_{\alpha}^2 
(V_{3q}^* V_{33})^2 + {\cal O} \left( \frac{m_q}{m_b}\right)~, \ee
where we used the vacuum saturation approximation multiplied by the bag 
factor $B_{2q}$
$$ <B_q^0|(\bar{q}_Lb_R)^2|\bar{B}^0>=
-\frac{5}{24}m_{B_q}f_{B_q}^2 B_{2q}\left(\frac{m_{B_q}}{m_b+m_q}\right)^2 
~.$$
The ${\cal O} \left( \frac{m_q}{m_b}\right)$ refer to contributions 
coming from  the $F^{d*}_{3q}$ terms in eq.(\ref{fcnc}). Using the 
vacuum saturation approximation and eq.(\ref{fij2}), these terms are
estimated to be only a few \% of the first term in eq (\ref{m12qh0})
for $q=s$ and much smaller for $q=d$. 
We do not display here the QCD corrections to $(M_{12})^{H^0}$.
Such corrections can be significant and play important role in the 
precise determination of the SM parameters. In contrast, the above 
expressions contain several unknowns of the Higgs sector because of which 
we prefer to simplify the analysis and retain only the leading terms
as far as the Higgs contributions to various observables are concerned.
The SM contribution is given by 
\be\label{m12sm}
(M_{12}^q)^{SM}=\frac{G_F^2 m_W^2m_{B_q}f_{B_q}^2 B_{q}\eta_B}{12 \pi^2}
(V_{3q}^* V_{33})^2 S_0(x_t) ~, \ee
with $S_0(x_t)\approx 2.3$ for $m_t\approx 161$ GeV.
Eqs.(\ref{m12qh0},\ref{m12sm}) together imply
\be \label{kapa}
\kappa^q\equiv
\left|\frac{(M_{12}^q)^{H^0}}{(M_{12}^q)^{SM}}\right|=
\left(\frac{5\sqrt{2}\pi^2|C_{\alpha}|^2}{G_F 
M_W^2\sin^2 
2\theta}\right)\left(\frac{m_b}{M_\alpha}\right)^2
\frac{B_{2q}}{B_{B_d}\eta_B}\left(\frac{m_{B_q}}{m_b+m_q}\right)^2+
{\cal O} \left( \frac{m_q}{m_b}\right)~. 
\ee
The neutral Higgs contribution to $\epsilon$  is given by
\be\label{epsh0}
\epsilon^{H^0}=\frac{5 G_F m_{K}f_{K}^2 B_{2K}}{12 
\sin^2 
2\theta \Delta 
m_K M_\alpha^2}\left(\frac{m_{K}}{m_s+m_d}\right)^2 
Im(F_{12}^{d} C_\alpha)^2 ~, 
\ee
Using the expression of $F_{12}^d$ from eq.(\ref{fij2}) and the 
Wolfenstein parameterization, 
one can rewrite the above equation as
\be\label{epsh02}
\epsilon^{H^0}\approx\frac{5 G_F m_s^2 m_{K}f_{K}^2 B_{2K}}{12 
\sin^2 
2\theta \Delta 
m_K M_\alpha^2}\left(\frac{m_{K}}{m_s+m_d}\right)^2 |C_\alpha|^2 A^4 
\lambda^{10} 
[(1-\bar{\rho})^2+\bar{\eta}^2]^{1/2}\sin 2(\eta_\alpha-\beta) , 
\ee
where $\tan\beta=\frac{\bar{\eta}}{1-\bar{\rho}}$ is one of the angles of 
the 
unitarity triangle.  The Higgs contribution to $\epsilon$ is suppressed
here by the strange quark mass and $\epsilon^{H^0}$ is practically 
negligible compared to $\epsilon^{SM}$:
\be \label{epsh0byepssm}
|\frac{\epsilon^{H^0}}{\epsilon^{SM}}|\approx 3.8 10^{-4} 
\frac{B_{2K}}{B_K}~ 
\frac{|C_{\alpha}|^2}{\sin^2 2\theta} \left(\frac{100 {\rm 
GeV}}{M_\alpha}\right)^2~
\frac{\sin 2(\eta_\alpha-\beta)}{\cos\beta+0.1 \sin\beta}~. \ee
The neutral Higgs contribution to the $K^0-\bar{K}^0$ mass difference is 
even
more suppressed compared to its experimental value.
\subsection{Experimental Inputs}
Constraints on the present scheme come from several independent
measurements. The complex amplitude $M_{12}^d$ is known quite well.
The magnitude is given in terms of the $B^0_d-\bar{B}^0_d$ mass 
difference \cite{f1}:
\be \label{mbd} \Delta M^d\equiv 2|M_{12}^d|=(0.507\pm 0.005)~{\rm 
ps}^{-1} 
~.\ee
The phase $\phi_d$ is measured through the mixing induced CP asymmetry
in the $B_d^0\rightarrow J/\psi K_S$ decay:
\be \label{phid} \sin \phi_d=0.668\pm 0.028~.\ee
Likewise, the $B^0_s-\bar{B}^0_s$ mass difference is quite well 
determined:
\be \label{mbs} \Delta M^s\equiv 2|M_{12}^s|=17.77\pm 0.12~{\rm ps}^{-1} 
~.\ee 
The corresponding phase $\phi_s$ is determined \cite{nir}  by the $D0$ 
collaboration \cite{d0} 
\be \label{phis} \phi_s=-0.70^{+0.47}_{-0.39} ~.\ee
by combining their measurements of (1) the light and the heavy $B_s^0$ 
width difference (2) the time dependent angular distribution in the 
$B_s^0\rightarrow J/\psi \phi$ decay and (3) the semileptonic charge 
asymmetries in the $B^0$ decays.

The SM predictions for the above quantities depend on the
hadronic and the CKM matrix elements. The determination of 
$\bar{\rho},\bar{\eta}$ is somewhat non-trivial when new physics is 
present.
The conventional SM fits use the  loop induced variables 
$\epsilon,M_{12}^d,\phi_d$ for determining $\bar{\rho},\bar{\eta}$.
These variables are susceptible to new physics contributions. This makes
extraction of $\bar{\rho},\bar{\eta}$ model-dependent. It is still
possible to determine these parameters and construct a 
universal unitarity triangle \cite{uut} for a unitary $V$
by assuming that the tree level contributions in the SM are not 
significantly affected by new physics. 
In that case, one can use 
only the tree level measurements for determining $\bar{\rho},\bar{\eta}$ 
\cite{bf}. Alternatively one can allow for NP contributions 
\cite{utfit1,utfit2,ckmfit,effectivemfv,mfv,cmfv,nmfv} 
in the loop induced processes while determining elements of $V$. The 
tree level observables are the moduli of $V$ and the unitarity angle
$\gamma$ \cite{f1}.
\beqa \label{modulie}
\lambda=|V_{us}|=0.2258\pm 0.0014&,& A=\frac{|V_{cb}|}{\lambda^2}=0.82\pm 
0.014 
\nonumber ~,\\
|V_{ub}|^{{\rm excl.}}=0.0034\pm 0.0004&,&
|V_{ub}|^{{\rm incl.}}=0.0045\pm 0.0003 ~. \eeqa
$\gamma$ is determined 
from purely tree level decay $B\rightarrow D^* K^*$. We will use the UTfit 
average value \cite{utfit2}:
\be \label{gamma}
\gamma=(83\pm 19)^\circ~. \ee
In terms of the Wolfenstein parameters, 
\beqa \label{rb}
\bar{\rho}=R_b \cos \gamma&,& \bar{\eta}=R_b \sin \gamma ~, \nonumber \\
R_b\equiv 
(1-\frac{\lambda^2}{2})\frac{1}{\lambda}|\frac{V_{ub}}{V_{cb}}|&=& 0.46\pm 
0.03 ~~~~ {\rm inclusive ~determination} ~, \nonumber \\
&=& 0.35\pm 
0.04 ~~~~ {\rm exclusive~ determination}
~. \eeqa
Eqs.(\ref{gamma}) and (\ref{rb}) provide a NP independent determination
of  $\bar{\rho},\bar{\eta}$, {\it e.g.} with inclusive values in 
eq.(\ref{modulie}),   
\beqa\label{tree}
\bar{\rho}=&0.06\pm 0.15,& \bar{\eta}=0.46\pm 0.03 ~. \\ \eeqa
One could use the above values of $\bar{\rho},\bar{\eta}$ to obtain
predictions of  $\epsilon$ and $M_{12}^d$ in the SM.  
The errors involved are rather large but it has the advantage of being
independent of any new physics contributing to these observables. 
This approach has been used 
for example in \cite{bf,np2,utfit1} to argue that a non-trivial NP phase is 
required
if $|V_{ub}|$ is close to its inclusive determination. We will use 
an alternative analysis which also leads to the same conclusion. 
The new physics contributions to the loop induced 
$\Delta F=2$ observables is parameterized as follows
\beqa \label{nppara}
M_{12}^q&=& (M_{12}^q)^{SM}(1+\kappa_q e^{i \sigma_q})=\rho_q (M_{12}^q)^{SM}
e^{i\phi_q^{NP}}~,\nonumber 
\\
\epsilon&=&\rho_{\epsilon}~\epsilon_{SM} ~. \eeqa
Model independent studies using the above or equivalent parameterization 
have been used 
to determine $\bar{\rho},\bar{\eta},\kappa_q,\sigma_q,C_{\epsilon}$ in
number of different work\cite{utfit1,utfit2,ckmfit,bf}. We will use the 
results from UTfit group whenever appropriate.

In view of the several unknown Higgs parameters, we make a simplifying 
assumption that only one Higgs  
contributes dominantly. We distinguish two qualitatively different 
situations corresponding to the dominance of the charged Higgs $H^+$ or of
a neutral Higgs.

\subsection {Charged Higgs dominance:} 
The effects of the charged Higgs on the $P^0-\bar{P}^0$ mixing as well as 
on
$\Delta F=1$ processes such as $b\rightarrow s\gamma$ have been discussed
at length in the literature \cite{franzini,2Hdm1,2Hdm2,bsg}. 
The present case remains unchanged compared to the standard two Higgs 
doublet model of type II
as long as the down quark mass dependent terms are neglected in 
eq.(\ref{ffh+}). Just for illustrative purpose and completeness we discuss 
some of the restrictions on the charged Higgs couplings and masses in this 
subsection before turning to our new results on the  neutral Higgs 
contributions to flavour violations.

The allowed values of $\bar{\rho},\bar{\eta}$ in the presence of the 
charged Higgs follow from the 
detailed numerical fits in case of MFV scenario, e.g. fits in
\cite{utfit1} give
\beqa \label{paramh+}
\bar{\rho}=0.154\pm 0.032&,& \bar{\eta}=0.347\pm 0.018 ~. \eeqa
We can substitute these values in the SM expressions for $\Delta M^d$
and $\epsilon$ to obtain \cite{f1}
\beqa \label{rodroe}
\rho_d\equiv &\frac{\Delta M^d}{(\Delta M^d)^{SM}}&=0.99\pm 0.29 ~, 
\nonumber \\
\rho_{\epsilon}\equiv &\frac{\epsilon}{\epsilon^{SM}}&=0.94\pm 0.09 ~.
\eeqa
This can be translated into bounds on $M_{H^+}$ and 
$\tan\theta$ using eqs.(\ref{k+}, \ref{epsh+}) and eq.(\ref{epssm}).
The $2\sigma$ bounds following from eq.(\ref{rodroe}) are shown in 
Fig.(1). The constraints from $\epsilon$ are stronger and allow
the middle (dotted ) strip in the $M_{H^+}-\tan\theta$ plane. 
These are illustrative bounds and we refer to literature 
\cite{franzini,2Hdm1,2Hdm2,bsg} for more detailed 
results which include QCD corrections. Generally, there is sizable 
region in $\tan\theta$, $M_{H^+}$ plane (e.g. $\tan\theta\gsim 1-2$ in 
Fig.(1)) for which the top induced charged 
Higgs contribution to $\rho_{d,\epsilon}$ is not important.
But the neutral 
Higgs can contribute to these observables in these regions as we  
now discuss.
\begin{figure}[ht] \begin{center}
\includegraphics[height=7.6cm,width=0.45\textwidth,angle=0]{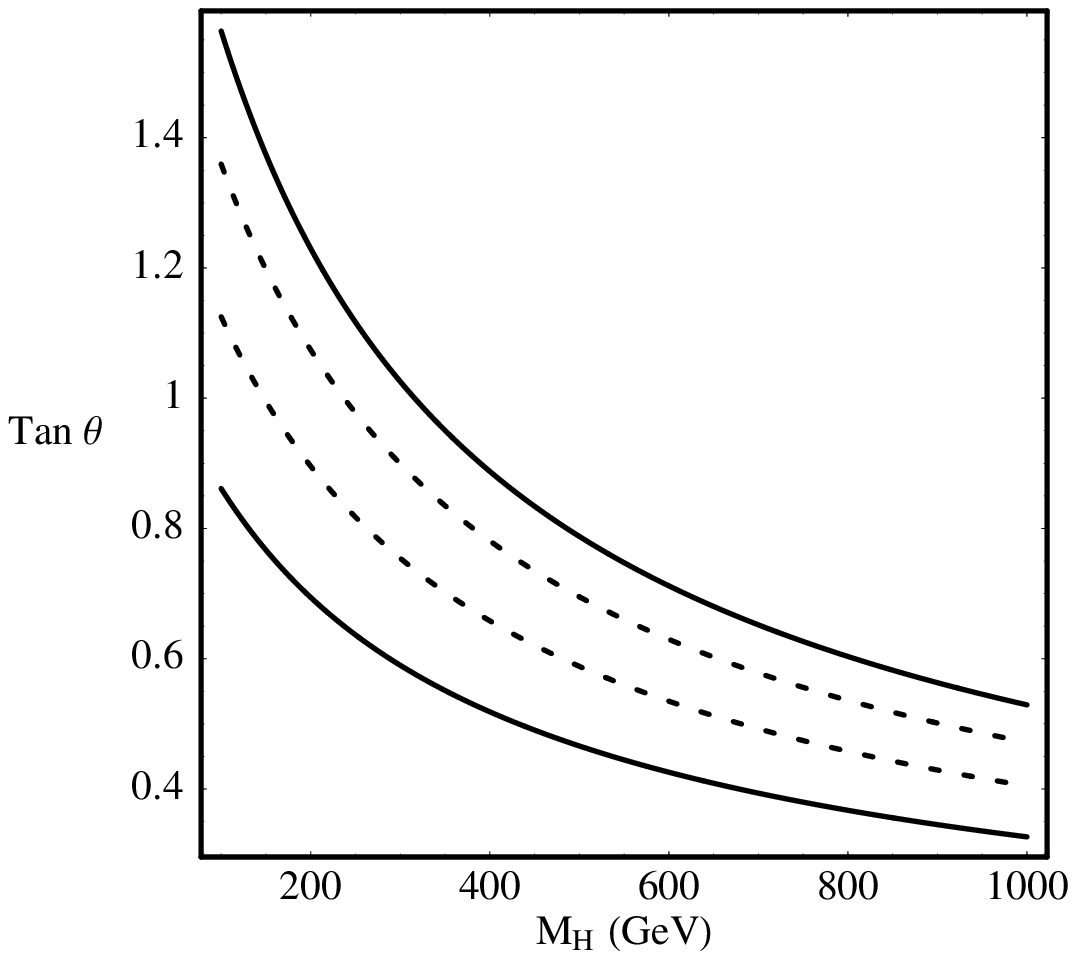}\hskip 
0.75cm 
\includegraphics[height=7.6cm,width=.45\textwidth,angle=0]{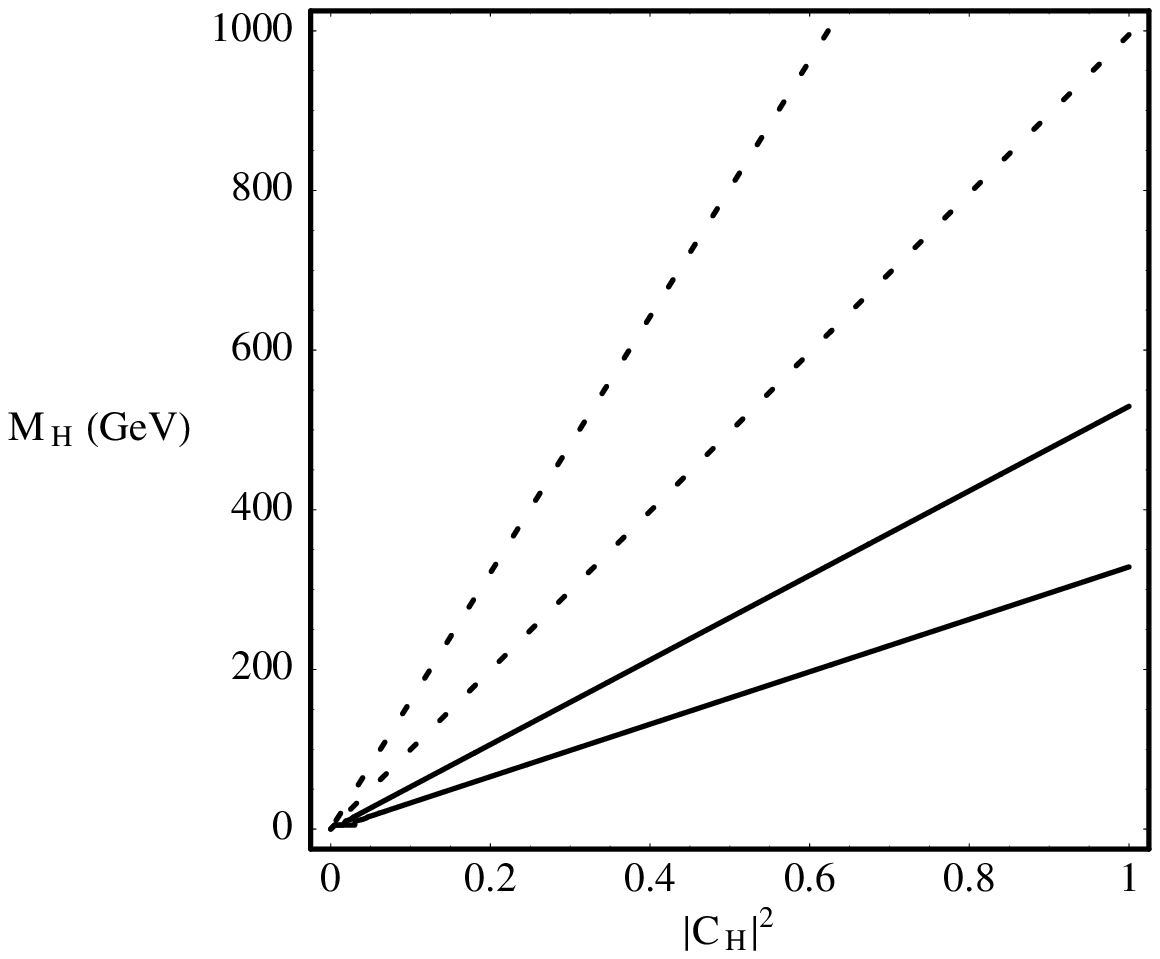}
\end{center}
\caption{Left panel: The 2$\sigma$ region  in the $\tan \theta,M_{H^+}$  
plane allowed by $\rho_d$ (solid) and $\rho_\epsilon$ (dotted) given in 
eq.(\ref{rodroe}). Right panel: Allowed regions in 
$|C_H|^2,M_H$ plane  
following from the inclusive determination of $|V_{ub}|$ for 
$\tan\theta=3$ (solid) and 10 (dotted). The left (right) 
panel is based on the assumption that the charged Higgs 
(neutral Higgs) alone accounts for the required new physics contribution
to $M_{12}^q$.} 
\end{figure}
\subsection{Neutral Higgs dominance:}
We label the dominating neutral Higgs field by 
$\alpha=H$ and retain only one term in eq.(\ref{m12qh0}). 
Unlike in the previous case, the neutral Higgs contribution to $\epsilon$
(and the $K^0-\bar{K}^0$ mass difference) is very small. It can 
contribute significantly to $M_{12}^{d,s}$ but these contributions are 
strongly correlated. Using eq.(\ref{m12qh0},\ref{kapa}) one finds that:
\beqa \label{ratio}
r=\frac{\kappa_s}{\kappa_d}&=&  
\frac{B_{2s}}{B_{2d}}
\frac{B_{B_d}}{B_{B_s}}\left(
\frac{m_{B_s}}{m_s+m_b}\right)^2 \left(
\frac{m_d+m_b}{m_{B_d}}\right)^2 ~,\nonumber \\ 
\sigma_d&=&\sigma_s=2 \eta_H ~.\eeqa
This ratio does not involve most of the unknown parameters and is 
determined by masses and the bag parameters. The ratios of 
$B$ parameter in eq.(\ref{ratio}) and hence $r$ is very close to 1.
For example, the results in \cite{lattice} for the bag parameters imply
\be\label{r}
r=1.04\pm 0.12 ~. \ee
Assuming $r=1$ leads to an important prediction:
$$\frac{\Delta M^s}{\Delta M^d}=\left(\frac{\Delta M^s}{\Delta 
M_d}\right)^{SM}~.$$
This prediction holds good in  various MFV scenario, e.g. SUSY MFV model  
at low $\tan \beta$ \cite{mfv}. Here it remains 
true even in the presence of an extra phase $\eta_H$.
The above  prediction can be usefully exploited \cite{uut} for the 
determination of 
one of 
the sides of the unitarity triangle:
\beqa \label{rt}
R_t&\equiv& 
\sqrt{(1-\bar{\rho})^2+\bar{\eta}^2}=\frac{1}{\lambda}|\frac{V_{td}}{V_{cb}}|~,\nonumber 
\\
&=&\frac{\xi}{\lambda}
\sqrt{\frac{M_{B_s}}{M_{B_d}}}\sqrt{|\frac{\Delta M_{B_d}}{\Delta 
M_{B_s}}|}\approx ~0.93\pm 0.05, \eeqa
where $\xi=\frac{f_{B_s}^2 B_{B_s}}{f_{B_d}^2 B_{B_d}}=1.23\pm 0.06$ 
\cite{utfit2}. We 
used the SM expression, eq.(\ref{m12sm}) in the above 
equation and the approximation $|V_{ts}|=|V_{cb}|$. 

The SM prediction for $\Delta M_s$ is independent of 
$\bar{\rho},\bar{\eta}$. Using, $f_{B_s}\sqrt{B_s}=0.262\pm 0.035$ MeV 
\cite{utfit2} we obtain
\be \label{rhos}
\rho_s\equiv \left|\frac{\Delta M_s}{\Delta M_s^{SM}}\right|\approx 
~0.96\pm 0.26  ~.\ee

The existing fits to the $\Delta F=2$ processes in the presence 
of NP are carried out
in the context of the MFV \cite{effectivemfv,mfv,utfit1,utfit2} or  
NMFV 
\cite{nmfv} scenario or in a model independent 
manner \cite{utfit1,utfit2}. Most of these assume that NP contributes 
significantly to $\Delta S=2$
transition, particularly to $\epsilon$. This is not  the case here.
On the other hand the model independent fits neglect correlations
between $\Delta M_d,\Delta M_s$ as present here. In view of this,
we performed our own  but simplistic fits in the present case.
We use $\phi_d,\gamma,R_b,R_t, \rho_s$ and $\epsilon$ in the fits 
assuming all errors to be Gaussian. 
The
expressions and the experimental values for these quantities are
already given in respective equations. We use the 
standard model expression for $\epsilon$. 
We have used $r=1$ in eq.(\ref{ratio}) giving 
eq.(\ref{rt}) and $\rho_d=\rho_s\equiv 
\tilde{\rho}$ and $\sigma_d=\sigma_s\equiv\sigma$. The above six 
observables are 
fitted in terms of the four unknowns 
$\bar{\rho},\bar{\eta},\tilde{\rho},\phi_d^{NP}$. 
The fitted 
values of the parameters  are 
sensitive 
to  $|V_{ub}|$. The accompanying 
table contains
values of the fitted parameters and $1\sigma$ errors obtained in three 
cases which use (a) inclusive (b) exclusive and (c) average value of 
$|V_{ub}|$
as quoted in \cite{pdg}. 
The predictions based on the average values agree 
within $1\sigma$ with the corresponding detailed model independent fits
by the Utfit group \cite{utfit2}:  
$\bar{\rho}=0.167\pm 0.051~,\bar{\eta}=0.386\pm 0.035$.
The values of $\bar{\rho},\bar{\eta}$ in the fit directly 
determine the phase of $(M_{12}^d)^{SM}$:
$$ \sin 2 \beta_d=  
\frac{\bar{\eta}(1-\bar{\rho})}{\sqrt{\bar{\eta}^2+(1-\bar{\rho})^2}}~.$$
The phase $\phi_d$ as measured through $S(\psi K_S)$ is  then given by 
$$\phi_d=2 \beta_d+\phi_d^{NP}~,$$
where $\phi_d^{NP}$ is defined in  eq.(\ref{nppara}) and can also be 
written as
\be\label{phiqnp}
\tan\phi_q^{NP}=\frac{\kappa_q\sin\sigma_q}{1+\kappa_q \cos \sigma_q} ~. 
\ee
Results in the table imply that if  $|V_{ub}|$ is close to the 
exclusive value then the present results
are consistent with SM.
If $V_{ub}$  is 
large and close 
to the inclusive value then $\phi_d^{NP}$ is non-zero
at 2$\sigma$ level. This conclusion is similar to observations made 
\cite{bf} on the 
basis of the use of $R_b,\gamma$ alone but with somewhat different
input values then used here. 
A non-zero $\phi_d^{NP}$ (and hence $\sigma$) has important 
qualitative implication for the model under consideration. Non-zero 
$\sigma$ requires CP violating phase   
$\eta_H$ from the scalar-pseudoscalar mixing.
As already remarked the minimal 2HDM with symmetry as in 
(\ref{symmetry}) cannot lead to such a phase and more general model
with an additional singlet field will be required. Also the charged Higgs 
contribution by itself cannot account for such a phase.

At the quantitative level,  $\tilde{\rho}\not=1$ implies
restrictions on the Higgs parameters, $M_H,|C_H|,\theta$. These parameters 
are simply related to $\kappa\equiv|\tilde{\rho} e^{i\phi_d^{NP}}-1|$ 
which is 
related to the said parameters through eq.(\ref{kapa}). Results in
table imply $\kappa=0.18\pm 0.08$ if $|V_{ub}|=|V_{ub}^{incl}|$. 
The values of  $M_H$ and $|C_H|^2$ which reproduce this $\kappa$ within
1$\sigma$ range is shown in 
Fig.(1) for two illustrative values of
$\tan\theta=3,10$. Both these values of $\tan\theta$ are chosen to make 
the 
charged Higgs contribution to $\kappa$ very small.  Unlike general  models 
with
FCNC, relatively light Higgs is a possibility within the present scheme
and there exist large ranges in $\theta$ and $C_H$ which allow this. 
 
One major prediction of the model is equality of new physics contributions
to CP violation in the $B_d$ and $B_s$ system. If the top induced charged 
Higgs 
contribution dominates then this CP violation is zero. In the case of the
neutral Higgs dominance, the phases $\sigma_d$ and $\sigma_s$ induced by 
the Higgs mixing are equal see, eq.(\ref{ratio}). Since the ratio $r$ in 
this equation is 
nearly one, 
let us write $r=1+\delta_r$ with $\delta_r\approx \pm {\cal O} (0.1)$.
Then  $\phi_s^{NP}$ in eq.(\ref{phiqnp}) can be approximated as
\beqa \label{pspd}
\tan\phi_s^{NP}&\approx& \tan \phi_d^{NP} \left[1+\delta_r(1- \cot \sigma 
\tan 
\phi_d^{NP})\right] ~, \nonumber \\
&\approx& (1+\delta_r) \tan \phi_d^{NP} ~. \eeqa
This prediction is independent of the details of the Higgs parameters.
Its important follows from the fact the standard
CP phase in the $B_s$ system is quite small, $\beta_s\sim -1.0 ^\circ$.
Thus observation of a relatively large $\phi_s=2 \beta_s+\phi_s^{NP}$ will 
signal new physics. The predicted values of 
$\tan \phi_s$ based on eq.(\ref{pspd})
and the numerical values given in table give
\beqa \label{phisnp}
\tan\phi_s&\approx&-0.18\pm0.08~~~~~~{\rm inclusive} ~,\nonumber \\
&\approx& ~~.03\pm0.08~~~~~~~{\rm exclusive} ~,\nonumber \\
&\approx& -0.14\pm0.08~~~~~~{\rm average} ~.\eeqa
All these values are at present consistent with the experimental 
determination eq.(\ref{phis}), by the $D0$ collaboration \cite{d0}.
Significant improvements in the errors is foreseen in future at 
LHCb \cite{lhcb} and relatively  large $\phi_s$ following from the 
inclusive $V_{ub}$
can be seen. The above predictions show correlation with $V_{ub}$ and 
also with the CP violating phase $\phi_d$. So combined improved 
measurements of all three will significantly test the model. The 
predictions of $\phi_s$ in the present case are significantly different 
from several other new physics scenario allowing relatively large values
for $\phi_s$ \cite{largephis}.
 
\begin{table}  \centering
\begin{math}
\begin{array}{|c|c|c|c|} \hline
&|V_{ub}^{{\rm incl.}}|&|V_{ub}^{{\rm 
excl.}}|&|V_{ub}^{{\rm average}}| \\
\hline
\bar{\rho}&0.200\pm0.039&0.121\pm 0.042&0.186\pm 0.039\\
\bar{\eta}&0.391\pm 0.028&0.320\pm 0.026&0.378\pm 0.027 \\
\rho_{d,s}&0.96\pm 0.26&0.96\pm 0.26&0.96\pm0.26\\
\sin\phi_d^{NP}&-0.18\pm 0.08&0.03\pm 0.08&-0.14\pm .08\\
\hline
\hline
\end{array}
\end{math}
\caption{Determination of NP parameters and $\bar{\rho},\bar{\eta}$ from 
detailed fits to predictions of the neutral Higgs induced FCNC. See, text 
for more details}
\label{tab:fit}
\end{table}

\section{Summary}
The general two Higgs doublet models are theoretically disfavored because
of the appearance of uncontrolled FCNC induced through Higgs exchanges at 
tree level. We have discussed here the phenomenological implications 
of a particular class of models in which FCNC are determined in terms of the
elements of the CKM matrix. This feature makes these models  predictive 
and we have worked out major predictions of the scheme. 
Salient aspects of the scheme discussed here are
\begin{itemize}
\item Many of the predictions of the scheme are similar to various other 
models \cite{mfv} which display MFV. The tree level FCNC couplings 
are governed by the CKM elements and the down quark masses while the 
dominant part of the charged Higgs couplings involve the same CKM factors 
but the top quark mass. Both contributions can be important and there 
exists regions of parameters ($\tan\theta\gsim 2-3$) in which the former 
contribution dominates. Unlike general FCNC models, the neutral Higgs
mass as light as the current experimental bound on the SM Higgs is 
consistent with the restrictions from the $P^0-\bar{P}^0$ mixing, see 
Fig.(1).
\item The neutral Higgs coupling to $\epsilon$ parameter is suppressed
in the model by the strange quark mass. This prediction differs from the 
general MFV models where the top quark contributes equally to the 
$B^0-\bar{B}^0$  mixing and $\epsilon$. Detailed fits to experimental
data is carried out which determine the CKM parameters 
$\bar{\rho},\bar{\eta}$ as displayed in the table.
\item Noteworthy and verifiable prediction of the model is correlation
(eq. (\ref{pspd})) between the CP violation in $B_s-\bar{B}_s$, 
$B_d-\bar{B}_d$ systems and $|V_{ub}|$ as displayed in the table.
\item We have restricted ourselves to study of the the $\Delta F=2$ 
flavour violations in this paper. The tree level FCNC would give rise to 
additional contributions to $\Delta F=1$ processes and to new 
processes such as flavour changing neutral Higgs decays \cite{rishi}. 
Already existing information on the
$\Delta F=1$ and $\Delta F=2$ processes can be very useful in identifying
allowed parameter space and verifiable signatures of the model.
Such a study will be taken up separately.
\end{itemize} 
\vskip 1.0truecm
\noindent{\bf Acknowledgments:}\\[.25cm]
 We thank Namit Mahajan for 
discussions and 
useful suggestions.


\begin{thebibliography}{99}
\bibitem{np1} A.~Lenz and U.~Nierste,
  %``Theoretical update of B/s - anti-B/s mixing,''
  arXiv:hep-ph/0612167;
B.~Dutta and Y.~Mimura,
  %``Modification of the unitarity relation for sin(2beta)-V(ub) in
  %supersymmetric models,''
  Phys.\ Rev.\  D {\bf 75}, 015006 (2007)
  [arXiv:hep-ph/0611268];  
 U.~Nierste,
  %``B/d and B/s mixing: Mass and width differences and CP violation,''
  arXiv:hep-ph/0612310; P.~Ball,
  %``Constraints on new physics from gamma and |V(ub)|,''
  arXiv:hep-ph/0612325; F.~J.~Botella, G.~C.~Branco and M.~Nebot,
  %``CP violation and limits on new physics including recent B/s 
%measurements,''
Nucl.\ Phys.\  B {\bf 768}, 1 (2007)[arXiv:hep-ph/0608100];
Z.~Ligeti, M.~Papucci and G.~Perez,
  %``Implications of the measurement of the B/s0 - anti-B/s0 mass difference,''
  Phys.\ Rev.\ Lett.\  {\bf 97}, 101801 (2006)
  [arXiv:hep-ph/0604112].
%
\bibitem{bf} P.~Ball and R.~Fleischer,
  %``Probing new physics through B mixing: Status, benchmarks and 
%prospects,''
  Eur.\ Phys.\ J.\  C {\bf 48}, 413 (2006) [arXiv:hep-ph/0604249];
P.~Ball,
  %``Probing new physics through Bs mixing,''
  arXiv:hep-ph/0703214.
%
\bibitem{np2} A.~J.~Buras, R.~Fleischer, S.~Recksiegel and F.~Schwab,
  %``New aspects of B --> pi pi, pi K and their implications for rare 
%decays,'' 
Eur.\ Phys.\ J.\  C {\bf 45}, 701 (2006)[arXiv:hep-ph/0512032]

%
\bibitem{utfit1} M.~Bona {\it et al.}  [UTfit Collaboration],
  %``The UTfit collaboration report on the unitarity triangle beyond the
  %standard model: Spring 2006,''
  Phys.\ Rev.\ Lett.\  {\bf 97}, 151803 (2006)
  [arXiv:hep-ph/0605213], see also  http://www.utfit.org, UTfit Collaboration (M. Bona 
{\it et al}.)
%
\bibitem{utfit2}
 M.~Bona {\it et al.}  [UTfit Collaboration],
  %``Model-independent constraints on Delta F=2 operators and the scale of 
%New Physics,''
arXiv:0707.0636 [hep-ph].
%
\bibitem{ckmfit} See, H.~Lacker,
  %``CKM matrix fits including constraints on New Physics,''
  arXiv:0708.2731 [hep-ph] and   http://ckmfitter.in2p3.fr/ for the latest 
  references on
  the analysis by the CKMfitter group (J. Charles {\it et al}).
%
\bibitem{effectivemfv} G.~D'Ambrosio, G.~F.~Giudice, G.~Isidori and 
A.~Strumia,
  %``Minimal flavour violation: An effective field theory approach,''
  Nucl.\ Phys.\  B {\bf 645}, 155 (2002)
  [arXiv:hep-ph/0207036].
%
\bibitem{mfv} A.~Ali and D.~London,
  %``Profiles of the unitarity triangle and CP-violating phases in the  
  %standard model and supersymmetric theories,''
  Eur.\ Phys.\ J.\  C {\bf 9}, 687 (1999)
  [arXiv:hep-ph/9903535]; A.~J.~Buras,
  %``Minimal flavor violation,''
  Acta Phys.\ Polon.\  B {\bf 34}, 5615 (2003)
  [arXiv:hep-ph/0310208]; W.~Altmannshofer, A.~J.~Buras and D.~Guadagnoli,
  %``The MFV limit of the MSSM for low tan(beta): meson mixings revisited,'' 
arXiv:hep-ph/0703200; Y.~Nir,
  %``Probing new physics with flavor physics (and probing flavor physics 
%with new physics),''
  arXiv:0708.1872 [hep-ph]; Y.~Grossman, Y.~Nir, J.~Thaler, T.~Volansky 
and 
J.~Zupan,
  %``Probing Minimal Flavor Violation at the LHC,''
  arXiv:0706.1845 [hep-ph]; B.~Grinstein, V.~Cirigliano, G.~Isidori and 
M.~B.~Wise,
  %``Grand unification and the principle of minimal flavor violation,''
  Nucl.\ Phys.\  B {\bf 763}, 35 (2007)
  [arXiv:hep-ph/0608123]; B.~Grinstein,
  %``Minimal Flavor Violation,''
  arXiv:0706.4185 [hep-ph]. 
%
\bibitem{cmfv} M.~Blanke and A.~J.~Buras,
  %``Lower Bounds on Delta M_{s,d} from Constrained Minimal Flavour Violation,'
 JHEP {\bf 0705}, 061 (2007) [arXiv:hep-ph/0610037];
M.~Blanke, A.~J.~Buras, D.~Guadagnoli and C.~Tarantino,
  %``Minimal Flavour Violation Waiting for Precise Measurements of Delta 
%M_s,S_{psi phi}, A^s_SL, |V_ub|, gamma and B^0_{s,d} -> mu+ mu-,''
  JHEP {\bf 0610}, 003 (2006) [arXiv:hep-ph/0604057].
%
\bibitem{nmfv} K.~Agashe, M.~Papucci, G.~Perez and D.~Pirjol,
  %``Next to minimal flavor violation,''
  arXiv:hep-ph/0509117.
%
\bibitem{nfc} S. Weinberg, Phys.\ Rev.\ Lett.\  {\bf 37}, 657 (1976)%
%
\bibitem{franzini} L.~F.~Abbott, P.~Sikivie and M.~B.~Wise,
  %``Constraints On Charged Higgs Couplings,''
  Phys.\ Rev.\  D {\bf 21}, 1393 (1980); P.~J.~Franzini,
  %``B anti-B Mixing: A Review of Recent Progress,''
  Phys.\ Rept.\  {\bf 173}, 1 (1989).
%
\bibitem{nlo} J.~Urban, F.~Krauss, U.~Jentschura and G.~Soff,
  %``Next-to-leading order QCD corrections for the B0 anti-B0 mixing with 
%an extended Higgs sector,''
  Nucl.\ Phys.\  B {\bf 523}, 40 (1998)
  [arXiv:hep-ph/9710245].
%
\bibitem{2Hdm1} Z.~j.~Xiao and L.~Guo,
  %``B0 anti-B0 mixing and B --> X/s gamma decay in the third type 2HDM:
  %Effects of NLO QCD contributions,''
  Phys.\ Rev.\  D {\bf 69}, 014002 (2004)
  [arXiv:hep-ph/0309103];R.~A.~Diaz, R.~Martinez and C.~E.~Sandoval,
  %``Improving bounds on flavor changing vertices in the two Higgs doublet
  %model from B0 - anti-B0 mixing,''
  Eur.\ Phys.\ J.\  C {\bf 46}, 403 (2006)
  [arXiv:hep-ph/0509194].
%
\bibitem{2Hdm2} Y.~L.~Wu and Y.~F.~Zhou,
  %``F0 anti-F0 mixing and CP violation in the general two Higgs doublet
  %model,''
  Phys.\ Rev.\  D {\bf 61}, 096001 (2000)
  [arXiv:hep-ph/9906313].
%
\bibitem{mb}
  G.~C.~Branco and R.~N.~Mohapatra,
  %``Complex CKM from spontaneous CP violation without flavor changing  
%neutral %current,''
  Phys.\ Lett.\  B {\bf 643}, 115 (2006)
  [arXiv:hep-ph/0607271].
%
\bibitem{brancoreal} G. C. Branco, Phys. Rev. Lett. {\bf 44}, 504 (1980).
%
\bibitem{brancockm} F.~J.~Botella, G.~C.~Branco, M.~Nebot and 
M.~N.~Rebelo,
  %``New physics and evidence for a complex CKM,''
  Nucl.\ Phys.\  B {\bf 725}, 155 (2005)
  [arXiv:hep-ph/0502133].
%
\bibitem{bhavik} A.~S.~Joshipura and B.~P.~Kodrani,
  %``Complex CKM matrix, spontaneous CP violation and generalized   
%$\mu$-$\tau$ %symmetry,''
  arXiv:0706.0953 [hep-ph].
%
\bibitem{fcncm} D.~Atwood, L.~Reina and A.~Soni,
  %``Flavor changing neutral scalar currents at mu+ mu- colliders,''
  Phys.\ Rev.\ Lett.\  {\bf 75}, 3800 (1995)
  [arXiv:hep-ph/9507416],
  %``Probing flavor changing top - charm - scalar interactions in e+ e-
  %collisions,''
  Phys.\ Rev.\  D {\bf 53}, 1199 (1996)
  [arXiv:hep-ph/9506243],
  %``Phenomenology of two Higgs doublet models with flavor changing neutral
  %currents,''
  Phys.\ Rev.\  D {\bf 55}, 3156 (1997)
  [arXiv:hep-ph/9609279];E.~Lunghi and A.~Soni,
  %``Footprints of the Beyond in flavor physics: Possible role of the Top 
%Two Higgs Doublet Model,''
  arXiv:0707.0212 [hep-ph];
S.~L.~Chen, N.~G.~Deshpande, X.~G.~He, J.~Jiang and L.~H.~Tsai,
  %``Spontaneous CP violating phase as the CKM matrix phase,''
  arXiv:0705.0399 [hep-ph].
%
\bibitem{asj1} A.~S.~Joshipura,
  %``Neutral Higgs and CP violation,''
  Mod.\ Phys.\ Lett.\  A {\bf 6}, 1693 (1991).
%
\bibitem{asj2} A.~S.~Joshipura and 
S.~D.~Rindani,
  %``Naturally suppressed flavor violations in two Higgs doublet models,''
  Phys.\ Lett.\  B {\bf 260}, 149 (1991).
%
\bibitem{bgl} G.~C.~Branco, W.~Grimus and L.~Lavoura,
  %``Relating the scalar flavour changing neutral couplings to the CKM
  %matrix,''
  Phys.\ Lett.\  B {\bf 380}, 119 (1996) [arXiv:hep-ph/9601383].
%
\bibitem{hall} A.~Antaramian, L.~J.~Hall and A.~Rasin,
  %``Flavor changing interactions mediated by scalars at the weak scale,''
  Phys.\ Rev.\ Lett.\  {\bf 69}, 1871 (1992)
  [arXiv:hep-ph/9206205]; L.~J.~Hall and S.~Weinberg,
  %``Flavor changing scalar interactions,''
  Phys.\ Rev.\  D {\bf 48}, 979 (1993)[arXiv:hep-ph/9303241].
%
\bibitem{asj} Anjan S. Joshipura, Phys.\ Lett.\  B {\bf 126}, 325 (1983).
%
\bibitem{f1} Theoretical and experimental inputs used throughout this paper are from the latest UTfit analysis given \cite{utfit2}.
%
\bibitem{nir} Y.~Grossman, Y.~Nir and G.~Raz,
  %``Constraining the phase of B/s - anti-B/s mixing,''
  Phys.\ Rev.\ Lett.\  {\bf 97}, 151801 (2006)
  [arXiv:hep-ph/0605028].
\bibitem{d0} V.~M.~Abazov {\it et al.}  [D0 Collaboration],
  %``Combined D0 measurements constraining the CP-violating phase and 
%width difference in the B/s0 system,''
  arXiv:hep-ex/0702030.
%
\bibitem{uut} A.~J.~Buras, P.~Gambino, M.~Gorbahn, S.~Jager and 
L.~Silvestrini,
  %``Universal unitarity triangle and physics beyond the standard model,''
  Phys.\ Lett.\  B {\bf 500}, 161 (2001)
  [arXiv:hep-ph/0007085].
%
\bibitem{bsg}
F.~Borzumati and C.~Greub,
  %``Two Higgs doublet model predictions for anti-B --> X/s gamma in  NLO 
%{QCD}(Addendum),''
  Phys.\ Rev.\  D {\bf 59}, 057501 (1999)[arXiv:hep-ph/9809438];
M.~Misiak {\it et al.},
  %``The first estimate of B(anti-B --> X/s gamma) at O(alpha(s)**2),''
  Phys.\ Rev.\ Lett.\  {\bf 98}, 022002 (2007)
  [arXiv:hep-ph/0609232].
%
\bibitem{lattice} D.~Becirevic, V.~Gimenez, G.~Martinelli, M.~Papinutto 
and J.~Reyes,
  %``Combined relativistic and static analysis for all Delta(B) = 2
  %operators,''
  Nucl.\ Phys.\ Proc.\ Suppl.\  {\bf 106}, 385 (2002)
  [arXiv:hep-lat/0110117]. 
%
\bibitem{pdg} W.-M. Yao et al., J. Phys. G 33, 1 (2006).
%
\bibitem{lhcb} N.~Magini  [CMS Collaboration],
  %``B/s --> J/psi Phi LHC review,''
  Nucl.\ Phys.\ Proc.\ Suppl.\  {\bf 170}, 146 (2007).
%
\bibitem{largephis} %Large Time-dependent CP Violation in B_s^0 System and 
%Finite D0-D0bar Mass Difference in Four Generation Standard Mode.
Wei-Shu Hou, Makiko Nagashima, Andrea Soddu, arXiv:hep-ph/0610385;
W.~S.~Hou and N.~Mahajan,
  %``Comments on the A(SL)(s) and Delta(Gamma(s)) studies at the Tevatron 
and
  %beyond,''
  Phys.\ Rev.\  D {\bf 75}, 077501 (2007)
  [arXiv:hep-ph/0702163]. 
%
\bibitem{rishi} A.~Arhrib, D.~K.~Ghosh, O.~C.~W.~Kong and R.~D.~Vaidya,
  %``Flavor changing Higgs decays in supersymmetry with minimal flavor
  %violation,''
  Phys.\ Lett.\  B {\bf 647}, 36 (2007)
  [arXiv:hep-ph/0605056].
\end{thebibliography}
\end{document}